# Geometry, stability and thermal transport of hydrogenated graphene nanoquilts

Zhongwei Zhang[1], Yuee Xie[1†], Qing Peng[2], and Yuanping Chen[1*]

[1] Department of Physics, Xiangtan University, Xiangtan 411105, Hunan, P.R. China

[2] Department of Mechanical, Aerospace and Nuclear Engineering, Rensselaer Polytechnic Institute, Troy, NY 12180, U.S.A

**Abstract:** Geometry, stability, and thermal transport of graphene nanoquilts folded by hydrogenation are studied using molecular dynamics simulations. The hydrogenated graphene nanoquilts show increased thermodynamic stability and better transport properties than folded graphene structures without hydrogenation. For the two-fold graphene nanoquilt, both geometry and thermal conductivity are very sensitive to the adsorbed hydrogen chains, which is interpreted by the red-shift of PDOS. For the multi-fold nanoquilts, their thermal conductivities can be tuned from 100% to 15% of pristine graphene, by varying the periodic number or length. Our results demonstrated that the hydrogenated graphene nanoquilts are quite suitable to be thermal management devices.

# Ⅰ. Introduction

Graphene is a superior material for electronics and phononics [1-4] because of super high electron mobility [5] and thermal conductivity [6]. Various functionalizations [7-10] are used to tune physical properties of graphene in order to achieve desired devices. Folding [10-12] is considered as an effective way of functionalization. Although graphene is very hard in the planar direction, it can be easily warped in the out-of-plane direction. Folding can drastically change the physical properties of graphene [10-15]. For example, the Fermi velocity in the graphene is reduced by folding [13]; an armchair nanoribbon changes from semiconducting to metallic after folding [14]; the thermal conductivities of graphene decrease as folding occurs [15]. Based on these folded graphene structures, some devices have been proposed, such as resistors [16], diverters [16], and conductance modulators [17], etc.

Experimentally, folded graphene can be obtained by random ultrasonic stimulation [18] or patterned etched trenches [19] i.e., by external mechanical forces. However, the geometries of these folded structures are not able to be easily controlled [18,19] and they are not stable at high temperatures, especially for multi-folded graphene [15]. The former studies indicated that the physical properties of folded graphene are very sensitive to their geometries [10-15]. Therefore, another more effective way should be found to precisely fold graphene. In fact, advances in microscopy technologies have provided an exciting opportunity to manipulate graphene morphologies by chemical functionalization [20,21]. A recent study shows that graphene can be folded by hydrogenation with different folding angles [21,22]. These hydrogen folded structures are more suited for thermal devices, however, to our best knowledge thermal transport in these folded graphene, especially in the multi-folded graphene, has been reported scarcely.

In this paper, we study thermal transport properties of two-fold and multi-fold graphene by applying classical molecular dynamics methods. Because graphene can be folded like a quilt by hydrogenation, as shown in Figures 1(b) and 1(d), we call these folded graphene nanoquilts. The geometric evolution from pristine graphene to a two-fold nanoquilt is discussed at first. The thermal conductivities decrease rapidly with hydrogenation, ultimately to about 30% of pristine graphene. The tuning range of thermal conductivities is much larger than the case of folded graphene without hydrogenation. On the basis of the two-fold nanoquilt, thermal transports in multi-fold nanoquilts are studied. The multi-fold nanoquilts exhibit excellent thermal stability at high temperatures, and their thermal conductivities can be well modulated by periodic number and periodic length. We find that, after pristine graphene is folded into a multi-fold structure, its thermal conductivity can at most be reduced to 15%. These findings are very useful for the applications of graphene in thermal devices.

## II. Simulation method and model

Figure 1(a) shows a graphene nanoribbon with length $L_1$ and width $W$, where $n$ hydrogen chains are adsorbed on the middle of top surface. The optimized structure can be seen in Figure 1(b), showing the graphene nanoribbon being folded by hydrogenation. $\theta$ is used to represent the folding degree defined as the angle between two-side straight ribbons. The folding process likes a quilt being folded into two layers, so we call it a two-fold nanoquilt. Based on Figure 1(a), a periodic two-surface hydrogenated graphene nanoribbon is shown in Figure 1(c), in which $L_0$ represents the periodic length and $L_2$ is the total length. Figure 1(d) is the optimized structure, i.e., it forms a multi-fold graphene nanoribbon with a periodic number *m*, which is called multi-fold nanoquilts.

In our simulations, the *LAMMPS* package [23] with the Adaptive Intermolecular Reactive Empirical Bond Order (AIREBO) potential [24] is adopted. Periodic boundary conditions are applied only to the direction perpendicular to the functionalized rows and other two directions are set free. Then atomic positions are relaxed to obtain minimum energy configuration using the conjugate gradient algorithm. After relaxation, we extend the relaxed structure to the desired hydrogenated graphene nanoribbons. That is, we prepared the folded graphene nanoribbons by C-H chemical bonding effects.

In the non-equilibrium molecular dynamics simulation (NEMD), the AIREBO potential is still adopted. On each hydrogenated graphene nanoribbon, fixed boundary conditions are implemented with the atoms at the left and right ends are fixed at their equilibrium positions. Next to the boundaries, the adjacent two cells of atoms are coupled to Nosé–Hoover thermostats [25,26] with temperatures 320 K and 280 K, respectively. That is, the average temperature is 300 K with a temperature difference 20 K. From Fourier's law, the thermal conductivity $K$ is defined as

$$K = -\frac{J}{\nabla T \cdot S}, \qquad (1)$$

where $\nabla T$ is the temperature gradient in structure, $J$ is the heat flux from the heat bath to the system, which can be obtained via calculating the heat baths power. $S = W \times H$ is the cross-sectional area, $W$ is the width and $H$ is the thickness of graphene nanoribbons, which we have chosen as $H = 0.144$ nm. In order to focus on the hydrogenation and deformation effect on phonons transport in nanoribbons, we keep the cross section area as a constant. With a time step of 0.5 fs, the graphene ribbon is first relaxed to the equilibrium states at 300 K for up to 500 ps under the canonical ensemble (NVT). Next, a temperature gradient is achieved by thermostat bath. After reaching the non-equilibrium steady state, 100 ps been used to calculate the thermal conductivity. In addition, the adopted AIREBO potential included the LJ term [24], which can be used to describe the *van der Waals forces* between ribbon parts after deformation, as shown in Figures 1(b) and 1(d).

To understand the underlying mechanisms of phonon transport in hydrogenated graphene, the phonon density of states (PDOS) has been calculated. The PDOS is calculated from the Fourier transform of the velocity autocorrelation function [26]:

$$\text{PDOS}(\omega) = \frac{1}{\sqrt{2\pi}} \int e^{-i\omega t} \langle \sum_{j=1}^{N} v_j(t) v_j(0) \rangle dt \qquad (2)$$

where $v_j(0)$ is the average velocity vector of a particle *j* at initial time, $v_j(t)$ is its velocity at time *t*, and $\omega$ is the vibration wavenumber.

## III. Results and discussion

In Figure 2(a), the geometric evolution of two-folded graphene nanoquilts by hydrogenation is shown. The size of the graphene nanoribbon we considered is $L_1$ = 6.3 nm and $W$ = 2.21 nm with armchair edge. The nanoribbon will be bended gradually as hydrogen chains are adsorbed on the middle. This is because the hydrogenation of graphene leads to a transformation of sp2 to sp3 and then causes a local distortion. The more chains, the more the bending deformation accumulates.

Figure 2(a) displays the relation between the bending angle $\theta$ and the number $n$ of adsorbed hydrogen chains. As the number $n$ increases from 1 to 9, the bending angle drops fast from $140^0$ to $25^0$, indicating the graphene nanoribbon is folded more and more. On one hand, the two-side straight nanoribbons come together gradually, on the other hand the middle region of ribbon forms a half cylinder. In the range of $10 < n < 18$, the bending angle approaches $\theta = 0°$, i.e., the two-side straight nanoribbons are parallel. As $n > 18$, the two straight nanoribbons are folded into a half cylinder. Generally speaking, the two straight nanoribbons are considered as two thermal leads in the application. The two leads are hidden in the case of $n > 18$, therefore we do not show them in Figure 2(a) anymore and we do not consider this case in the following study of thermal transport. Seen in Figure 2(a), one can find that the geometries of two-fold graphene nanoquilts can be tuned by hydrogen coverage. However, our results are somewhat different with those in Ref. [22] in which the angle of folded graphene varies linearly with hydrogenation ratio.

The corresponding thermal transport properties of the two-fold nanoquilts are shown in Figure 2(b). Somewhat similar to the geometric variation, thermal conductivities of the nanoquilts decrease quickly as the chains number $n$ of hydrogenation increases from 1 to 7, and then approach a convergent value which is about 1/3 the thermal conductivity of a pristine nanoribbon. Apparently, the combined effect of deformation and hydrogenation result in a remarkable reduction of thermal conductivity. As the adsorbed chains are above 10, the thermal conductivities of the nanoquilts are insensitive to the hydrogenation coverage because of their steady geometries. The thermal conductivity of folded graphene without hydrogenation is also shown in Figure 2(b) for comparison (see the dashed line). These folded structures are obtained by relaxation of hydrogenated nanoquilts after the hydrogen atoms are removed. Some folded structures are shown in the inset of Figure 2(b). One can find that the thermal conductivities of these folded nanoribbons (without-H) are less sensitive to the geometric deformation than those of hydrogenated nanoribbons. In other words, thermal transport can be more effectively modified by hydrogenated nanoquilts. As demonstrated by Ouyang [17], folded graphene is an excellent thermal conductance modulator. Obviously, the hydrogenated folded graphenes are more effective to construct thermal modulators and resistors.

It is well known that the thermal transport properties are determined by phonon activities. To understand the phonon transport in the two-fold nanoquilts, in Figure 3(a), the PDOS of these nanoquilts are calculated. Compared with the PDOS of the pristine nanoribbon, the PDOS of hydrogenated nanoquilts have an additional phonon peak at frequencies about 103 THz, which corresponds to C−H bond vibrations. It has no contribution to the thermal conductivity because its frequency is too high. In the lower frequency region, one can find that the G-band peak (the higher frequency peak) of the pristine nanoribbon at frequencies about 65 THz shifts to a lower frequency as the number of hydrogen chains *n* increases. The detailed shift of the G-band is shown in Figure 3(b). The frequency of the G-band is very sensitive to the hydrogenation at small covered chains. The red shift of the G-band induces the acoustic phonons to shift to low frequencies and also leads to a reduction of the phonon group velocity ( $v = dw/dq$ ). Meanwhile, the phonon mean free path *l* is decreased by hydrogenation [28] while the heat capacity *C* is relatively insensitive to the shift of the G-band at room temperature [29]. Therefore, thermal conductivities *K* of nanoquilts decrease because $K = \sum Cvl$, from classical lattice thermal transport theory. In the bottom panel of Figure 3(a), the PDOS of folded nanoribbons without hydrogenation indicates that the physical folding has little effect on the G-band and thus the change of thermal conductivity is also small.

The multi-folded graphene nanoquilts can be constructed based on the two-fold nanoquilt, as shown in Figure 1. Here, the thermal stability of multi-fold nanoquilts is discussed. We consider a nanoquilts consisting of 8 periodic two-fold nanoquilt with hydrogen chains $n = 5$ and $L_0 = 6.3$ nm. Figure 4(a) shows atomic structure of the nanoquilts at 0 K, while Figure 4(b) is the structure at 300 K. One can find that an increase in temperature does not destroy the form of the quilt. Our further calculations demonstrate that the folded nanoquilt maintains periodic structure up to temperature 1200 K (not shown). The main reason is the adsorbed hydrogen atoms protect the folding edges of the quilt. For comparison, Figures 4(c) and 4(d) display the folded graphene without hydrogenation at 0 K and 300 K, respectively. The periodic folded structure is destroyed obviously by the thermodynamic process. This indicates that the nanoquilts after hydrogenation are more stable than those obtained by external mechanical forces. Other nanoquilts with different periodic number $m$ in Figures 4(e) and 4(f) further show the stability of the hydrogenated nanoquilts. In addition, our calculations demonstrate that the multi-fold hydrogenated nanoquilts can maintain their geometric form until 1200 K. So, the hydrogenated folded graphene are more suitable for thermal devices.

Thermal transport in nanostructures is tightly associated with the size of the structures. In Figure 5, thermal conductivities of multi-fold nanoquilts as a function of periodic number $m$ are calculated. As the periodic number $m$ is small ($m < 10$), the thermal conductivities of all nanoquilts increase with their length, which is different with the results in Ref. [22]. As $m$ becomes larger, however, different kinds of multi-fold nanoquilts show different dependent of thermal conductivities on the length. For the multi-fold nanoquilts with $n$ = 1 and 3, the thermal conductivities continue to increase. For the multi-fold nanoquilts with $n$ = 5 and 7, their thermal conductivities approach converged values. The difference originates from the variation of transport mechanisms in the different kinds of nanoquilts. As the periodic number is small, the length of all nanoquilts is in the range of the phonon mean free path, i.e., phonon transport is ballistic. Because the thermal conductivity is proportional to the length of the ballistic mechanism, thermal conductivities of all nanoquilts increase in proportion to their length ($K \sim L^{\lambda}$, where $\lambda$ is the divergence power) [30]. The phonon mean free paths of the nanoquilts with more hydrogen chains ($n$ = 5 and 7) decrease sharply because of larger deformations and scattering of hydrogen atoms. Therefore, phonon transport in these nanoquilts changes from ballistic to diffusive transport. In this case the thermal conductivities are not sensitive to the length anymore, in larger $m$, and thus converge to a steady value. For the longer multi-fold nanoquilts with less hydrogen chains ($n$ = 1 and 3), phonon transports in these structures are between ballistic and diffusive transport, because the effect of adsorbed hydrogen atoms is relatively week. As a result, their thermal conductivities continue increasing but the rate of the increasing speed becomes slower. It is noted that the thermal conductivity of graphene nanoribbons is significantly reduced by multiple folds. For example, the thermal conductivity of a nanoquilt with $n$ = 7 and $m$ = 18 is 41.35 W/mK, which only about 15% of the pristine nanoribbon (287.2 W/mK). In addition, the VDWs interaction can also affects the thermal transport of nanoquilts. For example, the thermal conductivity of a nanoquilt with $m$ = 9 and $n$ = 3 decreases 13.5% as the VDWs force is omitted.

In Figure 6, the effect of the periodic length $L_0$ of multi-fold nanoquilts on thermal conductivity is shown. The nanoquilts we studied here are fixed at $L_2$ = 88.2 nm. The decrease of $L_0$ leads to the increase of periodic number $m$. As $L_0$ changes from 88.2 to 0 nm, it represents the structural change from a pristine nanoribbon to multi-fold nanoquilts and then to a graphane nanoribbon. It is seen from Figure 6 that, thermal conductivities of all kinds of the nanoquilts ($n = 1, 3, 5$ and $7$) decrease with $L_0$, which is attributed to more and more periodic folds strengthening the phonon scattering. At $L_0 = 0$ nm, i.e., the structure is an unfolded graphane nanoribbon, the thermal conductivities undergo a small increase, 89.8 W/mK, which is about 46.5% of the pristine nanoribbon.

Finally, it should be noted that we also calculated the thermal transport in two-fold and multi-fold nanoquilts with zigzag edges. The results indicate that thermal transports in the zigzag-edge nanoquilts have the same properties as those in the armchair-edge nanoquilts, even though they have some differences in the detailed quantitative values because of anisotropy.

## IV. Conclusion

In summary, we have investigated thermal transport properties of two-fold and multi-fold graphene nanoquilts. The studies of the two-folded hydrogenated nanoquilts show that their bending angle and thermal conductivity depend on the adsorbed hydrogenated chains. The tuning range of thermal conductivity is much larger than the case of folded graphene without hydrogenation. The variation of thermal conductivity is interpreted by the red-shift of PDOS. The multi-fold hydrogenated nanoquilts are more stable than multi-folded graphene without hydrogenation at high temperatures. Moreover, the thermal conductivity of multi-folded graphene nanoquilts can be well modulated by varying periodic number $m$ and periodic length $L_0$. We have found that a thermal reduction of multi-folded graphene nanoquilts have been acquired as high as about 85% by varying periodic parameters. Our work demonstrates that hydrogenation is an effective method to precisely control geometry and also thermal transport of folded graphene. These findings may be useful for the applications of graphene in nanoscale devices, such as modulators, resistors and graphene electronic circuits.

**Acknowledgments**

This work was supported by the National Natural Science Foundation of China (Nos. 511 76161, 51376005 and, 11474243).

Corresponding author: † xieyech@xtu.edu.cn; * chenyp@xtu.edu.cn

**References**

[1] K. S. Novoselov, A. K. Geim, S. V. Morozov, D. Jiang, Y. Zhang, S. V. Dubonos, I. V. Grigorieva and A. A. Firsov, Science **306** (2004) 666.

[2] A. K. Geim and K. S. Novoselov, Nature Mater. **6** (2007) 183.

[3] A. H. C. Neto, F.Guinea, N. M. R. Peres, K. S. Novoselov and A. K. Geim, Rev. Mod. Phys. **81** (2009) 109.

[4] N. Yang, G. Zhang and B. Li, Appl. Phys. Lett. **95** (2009) 033107.

[5] C. R. Dean, A. F. Young, I . Meric, C. Lee, L. Wang, S. Sorgenfrei, K. Watanabe, T. Taniguchi, P. Kim, K. L. Shepard and J. Hone, Nature Nanotech. **5** (2010) 722.

[6] A. A. Balandin, S. Ghosh, W. Bao, I. Calizo, T. Teweldebrhan, F. Miao and C. N. Lau, Nano Lett. **8** (2008) 902.

[7] T. Ramanathan, A. A. Abdala, S. Stankovich,D. A. Dikin, M. H. Alonso, R. D. Piner, D. H. Adamson, H. C. Schniepp, X. Chen, R. S. Ruoff, S. T. Nguyen, I. A. Aksay, H. R. A. Prud and L. C. Brinson, Nature Nanotech. **3** (2008) 327.

[8] T. M. G. Mohiuddin, A. Lombardo, R. R. Nair, A. Bonetti, G. Savini, R. Jalil, N. Bonini, D. M. Basko, C. Galiotis, N. Marzari, K. S. Novoselov, A. K. Geim and A. C. Ferrari, Phys. Rev. B **79** (2009) 205433.

[9] D. Pan, J. Zhang, Z. Li and M. Wu, Adv. Mater. **22** (2010) 734.

[10] N. Patra, B. Wang and P. Král, Nano Lett. **9** (2009) 3766.

[11] S. T. Yang, Y. Chang, H. Wang, G. Liu, S. Chen Y. Wang, Y. Liu and A. Cao, J. Colloid Interface Sci. **351** (2010) 122.

[12] F. Liu, S. Song, D. Xue and H. Zhang, Adv. Mater. **24** (2012) 1089.

[13] Z. Ni, Y. Wang, T. Yu, Y. You and Z. Shen, Phys. Rev. B **77** (2008) 235403.

[14] Y. Xie, Y. Chen and J. Zhong, J. Appl. Phys. **106** (2009) 103714.

[15] N. Yang, X. Ni, J. W. Jiangand B. Li, Appl. Phys. Lett. **100** (2012) 093107.

[16] Y. Xie, Y. Chen, X. Wei and J. Zhong, Phys. Rev. B **86** (2012) 195426.

[17] T. Ouyang, Y. Chen, Y. Xie, G. M. Stocks and J. X. Zhong, Appl. Phys. Lett. **99** (2011) 233101.

[18] J. Zhang, J. Xiao, X. Meng, C. Monroe, Y. Huang, and J. M. Zuo, Phys. Rev. Lett. **104** (2010) 166805.

[19] K. Kim, Z. Lee, B. D. Malone, K.T. Chan, B. Alemán, W. Regan, W. Gannett, M. F. Crommie, M. L. Cohen, and A. Zettl, Phys. Rev. B **83** (2011) 245433.

[20] S. Akcöltekin, H. Bukowska, T. Peters, O. Osmani, I. Monnet, I. Alzaher, B. B. Etat, H. Lebius and M. Schleberger, Appl. Phys. Lett. **98** (2011) 103103.

[21] C. D. Reddy and Y. W. Zhang, Carbon **69** (2014) 86.

[22] C. Li, G. Li and H. Zhao, Carbon **72** (2014) 185.

[23] S. Plimpton, J. Comp. Phys. **117** (1995) 1.

[24] S. Stuart, A. Tutein and J. Harrison, J. Chem. Phys. **112** (2000) 6472.

[25] S. Nosé, J. Chem. Phys. **81** (1984) 511.

[26] W. G. Hoover, Phys. Rev. A **31** (1985) 1695.

[27] J. M. Dickey and A. Paskin, Phys. Re. **188** (1969) 1407.

[28] C. Li, G. Li and H. Zhao, Carbon **72** (2014) 185.

[29] M. G. Xia and S. L. Zhang, Eur. Phys. JB **84** (2011) 385.

[30] X. Mu, X. Wu, T. Zhang, D. B. Go and T. Luo, Sci. Rep. **4** (2014) 3909.

**Figure captions**

**Fig. 1.** (color online) (a) Schematic view of an armchair-edge graphene nanoribbon adsorbing $n$ hydrogen chains. $L_1$ and $W$ represent the length and width of nanoribbon, respectively. Top panel is the top view while the bottom panel is the side view. (b) Optimized structure of (a), a two-fold nanoquilt with a bending angle $\theta$. (c) Schematic view of a periodic hydrogenated graphene nanoribbon in units of (a). $L_2$ is the length of the nanoribbon and $L_0$ is the periodic length. (d) Optimized structure of (c), a multi-fold nanoquilts of $m$ period.

**Fig. 2**. (color online) (a) Bending angle $\theta$ of two-fold nanoquilts as a function of hydrogenated chains $n$. Several corresponding structures are shown in the inset. (b) Thermal conductivities of two-fold nanoquilts as a function of hydrogenated chains $n$. Thermal conductivities of folded graphene nanoribbons without hydrogenation are also shown for compare. The other parameters of the two-fold nanoquilt are $L_1 = 6.3$ nm and $W = 2.21$ nm.

**Fig. 3.** (color online) (a) Phonon PDOS for pristine graphene, two-fold hydrogenated nanoquilts ($n$ = 3, 5, 7 and 11), and folded graphene without hydrogenation (corresponding to the two-fold hydrogenated nanoquilt with $n$ = 11). The high frequency peaks represent the G-peak of phonon. (b) The G-peak for the two-fold nanoquilt as a function of hydrogenated chains $n$. The other parameters of the two-fold nanoquilt are $L_1 = 6.3$ nm and $W = 2.21$ nm.

**Fig. 4.** (color online) (a-b) Multi-fold nanoquilts with $n$ = 5 and $m$ = 8 under temperatures 0 K and 300 K, respectively. (c-d) Multi-fold nanoribbon without hydrogenation ($m$ = 8) under temperatures 0 K and 300 K, respectively. (e-f) Multi-folded nanoquilts ($n$ = 5) with periodic numbers $m$ = 4 and $m$ = 12 under 300 K,

respectively. The other parameters of the multi-fold nanoquilts are $L_0$ = 6.3 nm and $W$ = 2.21 nm.

**Fig. 5.** (color online) Thermal conductivities of multi-fold nanoquilts with different hydrogenated chains $n$ are as functions of periodic number $m$. The periodic length $L_0$ = 6.3 nm.

**Fig. 6.** (color online) Thermal conductivities of multi-fold nanoquilts with different hydrogenated chains $n$ are as functions of periodic length $L_0$. The total length of all nanoquilts are fixed at $L_2$ = 88.2 nm.

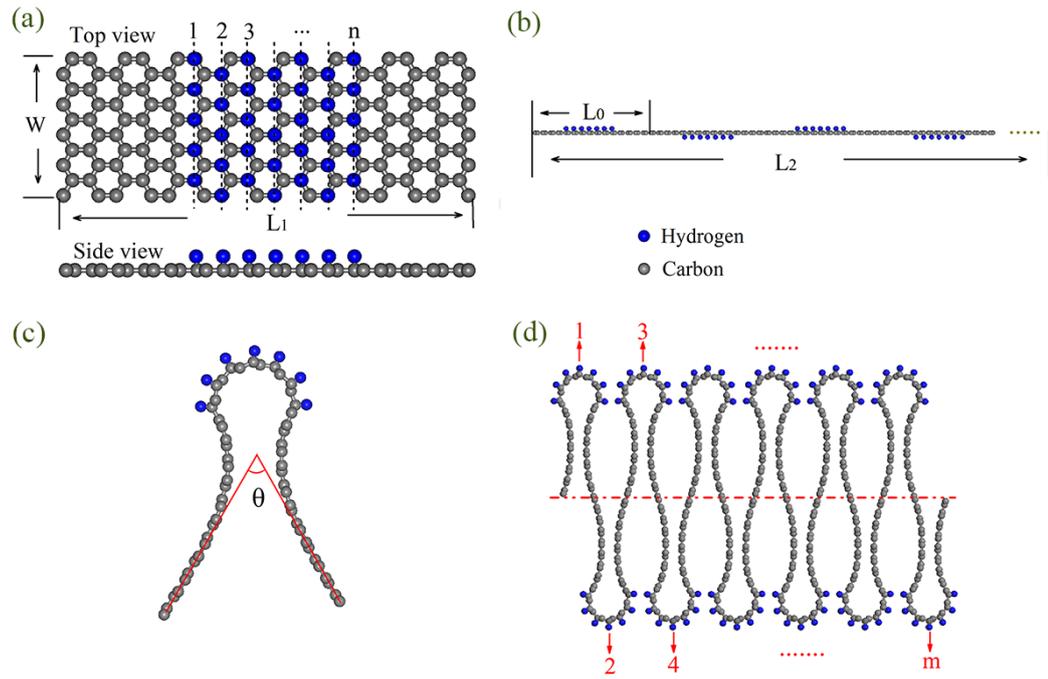

Fig. 1

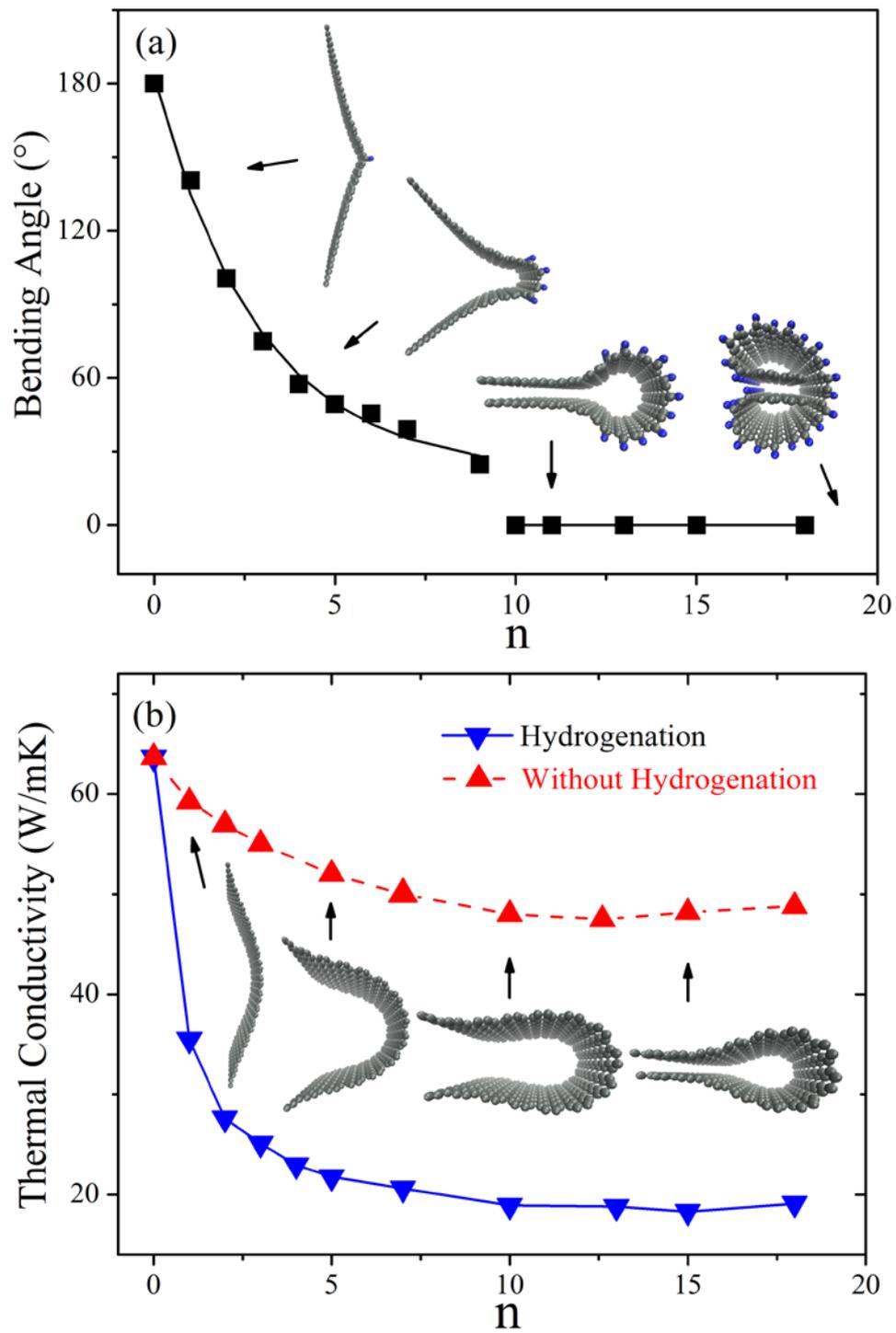

Fig. 2

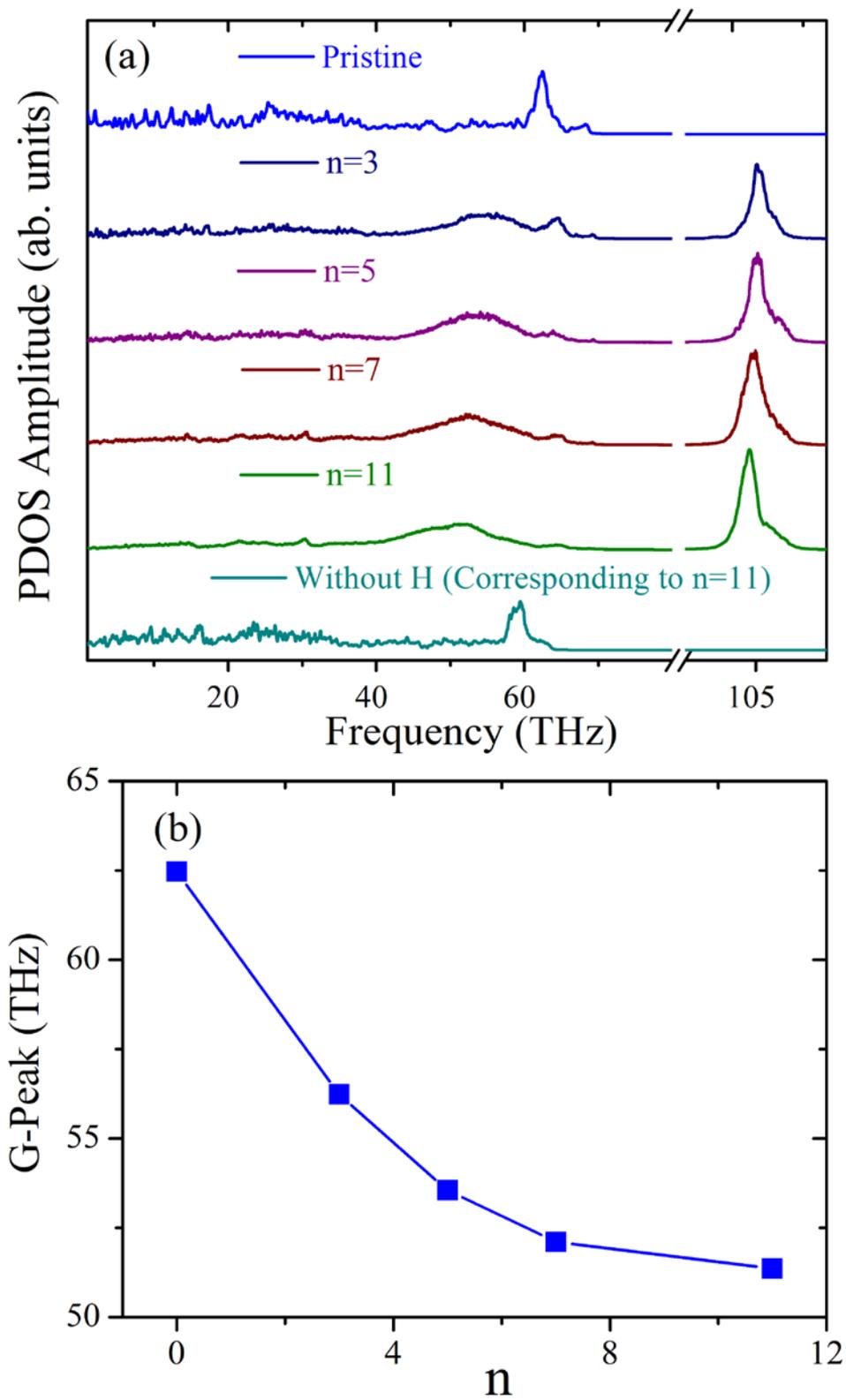

Fig. 3

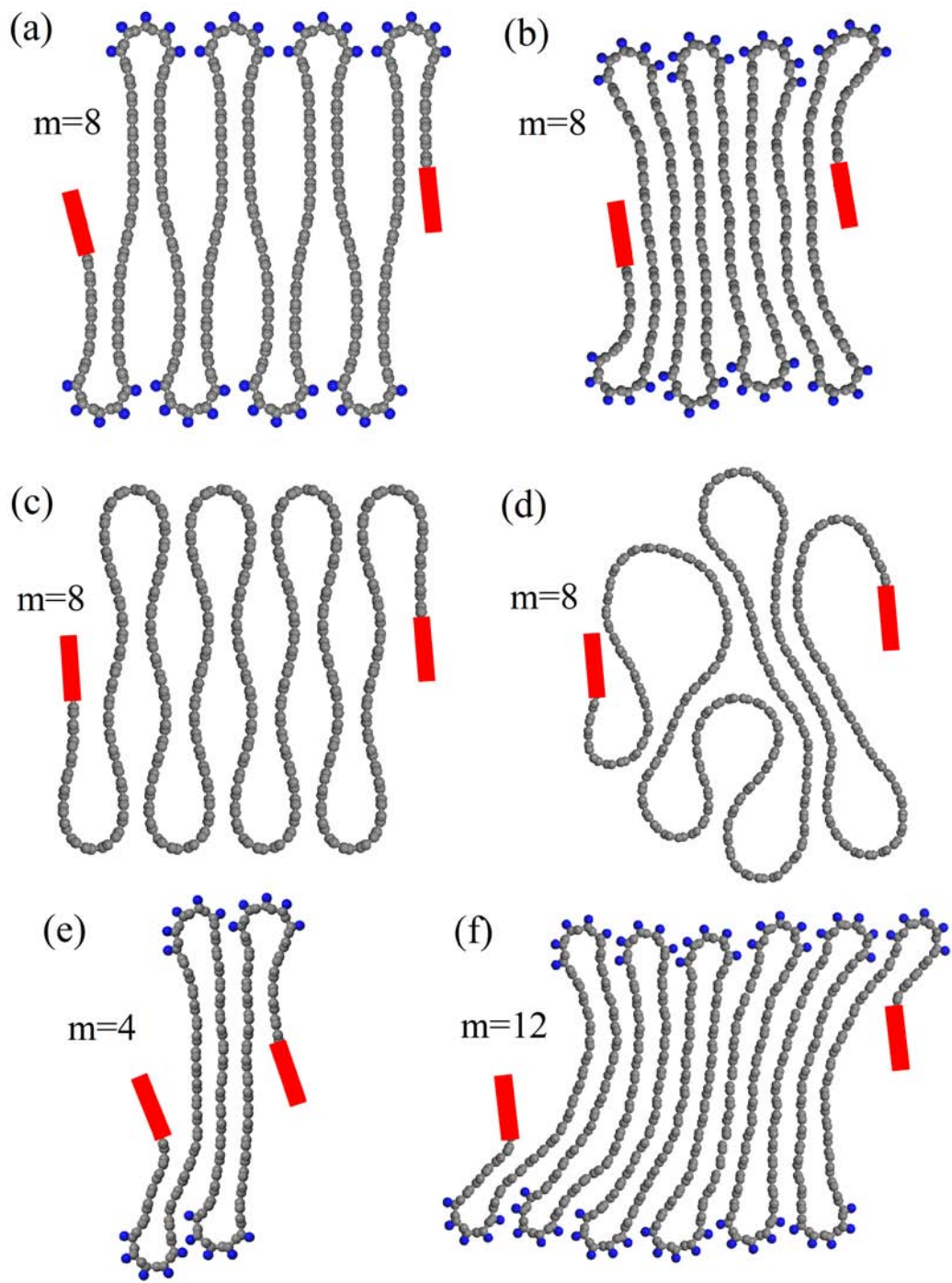

Fig. 4

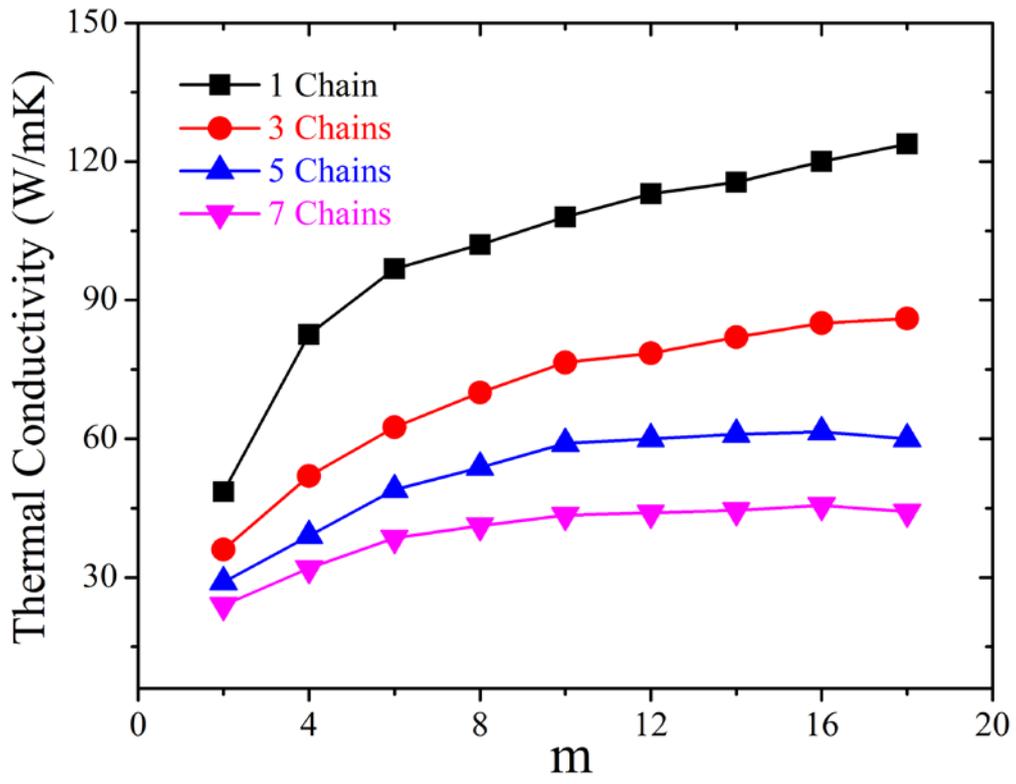

Fig. 5

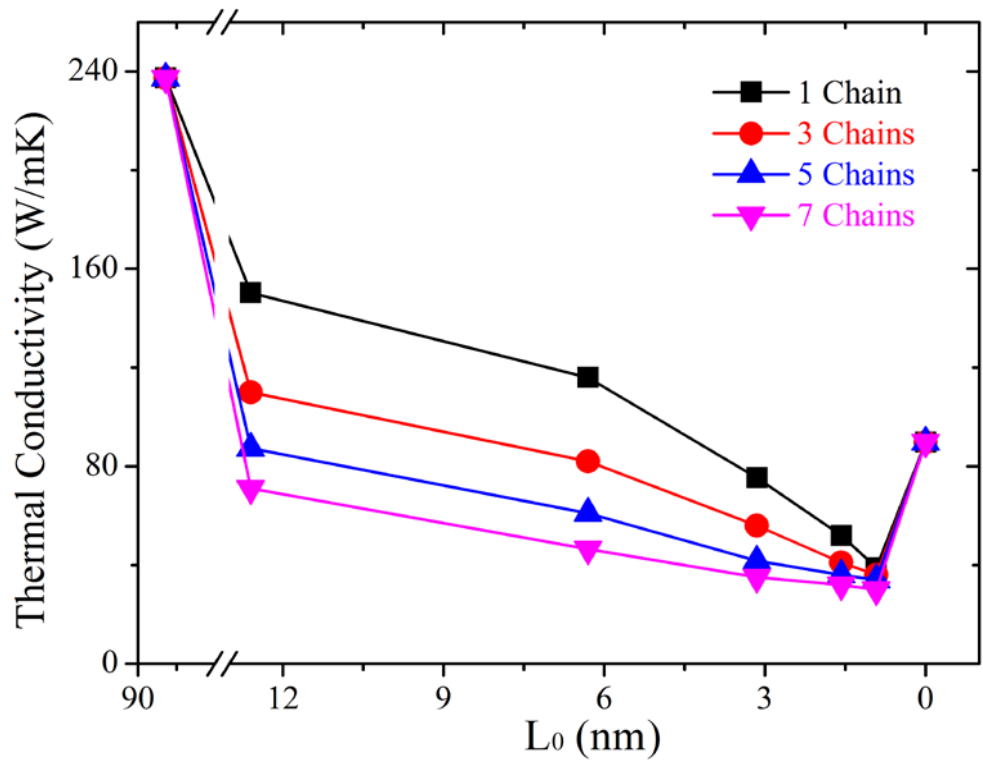

Fig. 6